\documentclass[12pt]{article}
\usepackage{graphicx}

\def\ignore#1{{}}

\newcounter{sxn}

\newcounter{axn}

\date{}

\newdimen\mybaselineskip
\newcommand{\beeq}{\begin{equation}}
\newcommand{\eneq}{\end{equation}}
\newcommand{\beqn}{\begin{eqnarray}}
\newcommand{\eeqn}{\end{eqnarray}}

\def\la{\raise.16ex\hbox{$\langle$}\lower.16ex\hbox{}  }
\def\ra{\, \raise.16ex\hbox{$\rangle$}\lower.16ex\hbox{} }

\def\psibar{ \psi \kern-.65em\raise.6em\hbox{$-$} \lower.6em\hbox{} }
\def\psibarb{ \psi \kern-.65em\raise.6em\hbox{$-$}  }

\begin{document}

\thispagestyle{empty}

\baselineskip=12pt



\vspace*{3.cm}

\begin{center}  
{\LARGE \bf  The Highly Real Quasinormal Modes of Schwarzschild-Anti De Sitter Black Holes}
\end{center}

\baselineskip=14pt

\vspace{2cm}
\begin{center}
{\bf  Ramin G. Daghigh}
\end{center}

\vspace{0.25 cm}
\centerline{\small \it Natural Sciences Department, Metropolitan State University, Saint Paul, Minnesota, USA 55106}
\vskip 0 cm
\centerline{}

\vspace{1cm}
\begin{abstract}
A recent investigation has led to the possibility of the existence of an interesting region of the asymptotic quasinormal mode spectrum of Schwarzschild-anti de Sitter black holes.  In this asymptotic region, the real part of quasinormal mode frequencies approaches infinity while the damping rate approaches a finite value.   These quasinormal mode frequencies were calculated using an analytic technique based on the complex coordinate WKB method.  In this paper, we use a different analytic technique to calculate such quasinormal mode frequencies.  The results of this paper provide further support of the possibility of the existence of these modes.

\baselineskip=20pt plus 1pt minus 1pt
\end{abstract}

\newpage
The asymptotic quasinormal mode (QNM) spectrum of four dimensional Schwarzschild-anti de Sitter (AdS) black holes were calculated analytically for the first time by Cardoso {\it et al} in \cite{Cardoso-N-S}, where the authors developed a method based on the monodromy technique of Motl and Neitzke\cite{Motl1, Motl2}.  Later, Natario and Schiappa\cite{Natario-S} generalized the analytic results of \cite{Cardoso-N-S} to spacetime dimensions greater than four.   In a recent paper\cite{Ramin-AdS}, Daghigh and Green reanalyzed the asymptotic QNM spectrum of Schwarzschild-AdS black holes in arbitrary spacetime dimensions ($D\ge 4$) using a different analytic technique called the complex coordinate WKB method\cite{Heading, Froman, Berry}.  This method was first applied to asymptotic QNMs of black holes by Andersson and Howls in \cite{Andersson}, where the authors calculate the asymptotic QNM frequencies of non-rotating black holes.  The authors of \cite{Ramin-AdS} not only confirm the analytic results found by Cardoso {\it et al}\cite{Cardoso-N-S} and Natario and Schiappa\cite{Natario-S}, but also demonstrate that in certain spacetime dimensions the analytic techniques they use predict the existence of new regions of the asymptotic QNM frequency spectrum.  These regions have not previously appeared or been explored in the literature\cite{Berti-K, Cardoso-K-L,Konoplya0, high-tone-1, high-tone-2, Ghosh}.  The QNMs of one of these regions, which appears only in even spacetime dimensions, resembles the normal modes in a pure AdS space.  The real part of the QNM frequencies of such a region approaches infinity while the damping rate approaches a finite value.  In \cite{Ramin-AdS}, the name ``highly real" is chosen for these modes to correspond to the widely used term ``highly damped", which is the situation when the imaginary part of the frequency is much larger than the real part.  If the highly real QNMs do indeed exist, it is argued in \cite{Ramin-AdS} that these modes become the most relevant modes in the context of AdS-CFT correspondence.  They can also serve as a testing ground for the new interpretation of the black hole asymptotic QNMs proposed by Maggiore\cite{Maggiore}.  The very interesting consequences of the existence of highly real QNMs motivates numerical and analytical investigations to prove or disprove the existence of these modes.  

\begin{figure}[tb]
\begin{center}
\includegraphics[height=5cm]{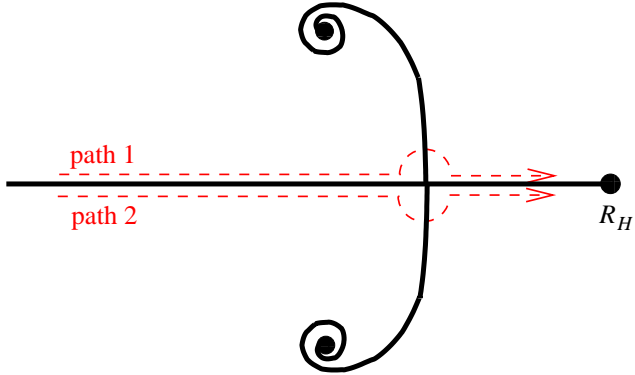}
\end{center}
\caption{Schematic presentation of the anti-Stokes lines in the complex $r$-plane for Schwarzschild-AdS black holes in four spacetime dimensions.  $R_H$ is the event horizon.  The two dashed arrows show the paths we take along anti-Stokes lines to determine the condition on the highly real QNM frequency.}
\label{Fig1}
\end{figure}

In this paper, we use the analytic technique which was developed by Cardoso {\it et al} in \cite{Cardoso-N-S} to see if we can reproduce the results of \cite{Ramin-AdS} for highly real QNMs.  At first glance, it is not obvious whether the two different analytic techniques mentioned above produce the same results.  Using the complex coordinate WKB method, Daghigh and Green\cite{Ramin-AdS} were able to obtain two unphysical and two physical solutions for highly real QNM frequencies by changing the dominancy of the WKB solutions on the pairs of Stokes lines, which are crossed on the path taken along anti-Stokes lines, and by rotating clockwise (in the upper half of the complex plane) and counter-clockwise (in the lower half of the complex plane) close to the singularity at the origin of the complex plane ($r=0$). 
In the method developed by Cardoso {\it et al}\cite{Cardoso-N-S}, there are no use of Stokes\footnote{Note that what we call anti-Stokes lines in this paper are called Stokes lines in \cite{Cardoso-N-S} and \cite{Natario-S}.} lines.  Therefore, one can only find two solutions for the highly real QNMs by rotating clockwise (path 1 in Fig. \ref{Fig1}) or counterclockwise (path 2 in Fig. \ref{Fig1}) close to the origin of the complex plane.  These two solutions could be the unphysical ones which were discarded by the authors of \cite{Ramin-AdS}.  If that is the case, then we can conclude that we have found strong evidence against the existence of the highly real QNMs.  In what follows, we derive an expression for highly real QNM frequencies using the analytic technique developed in \cite{Cardoso-N-S}.   

Various classes of non-rotating black hole metric perturbations in a spacetime with dimension $D\ge 4$ are governed generically by a Schr$\ddot{\mbox o}$dinger wave-like equation of the form
\beeq
{d^2\Psi \over dz^2}+\left\{ \omega^2-V\left[r(z)\right] \right\}\Psi =0 ~,
\label{Schrodinger}
\eneq
where $V(r)$ is the QNM potential obtained by Ishibashi and Kodama\cite{Ishibashi1, Ishibashi2, Ishibashi3} for scalar (reducing to polar at $D=4$), vector (reducing to axial at $D=4$), and tensor (non-existing at $D=4$) perturbations.  It is worthwhile to mention that the effective potential for tensor perturbations is equivalent to that of the decay of a test scalar field in a black hole background in every spacetime dimensions including $D=4$.  Following \cite{Cardoso-N-S} and \cite{Natario-S}, we assume the perturbations depend on time as $e^{i\omega t}$.  Consequently, in order to have damping, the imaginary part of $\omega$ ($\omega_I$)must be positive.  Keep in mind that in \cite{Ramin-AdS}, the authors assume the perturbations depend on time as $e^{-i\omega t}$ instead of $e^{i\omega t}$.   
In Eq.\ (\ref{Schrodinger}), the tortoise coordinate $z$ is defined by 
\beeq
dz ={dr \over f(r) }~,
\label{tortoise}
\eneq
where $f(r)$ is related to the spacetime geometry, and is given by 
\beeq
f(r)=1-{2\mu \over r^{D-3}}-\lambda r^2~.
\label{function f}
\eneq
The ADM mass, $M$, of the black hole is related to the parameter $\mu$ by
\beeq
M = {(D-2)A_{D-2} \over 8 \pi G_D}\mu~,
\eneq
where $G_D$ is the Newton gravitational constant in spacetime dimension $D$ and $A_n$ is the area of a unit $n$-sphere,
\beeq
A_n={2\pi^{n+1 \over 2} \over \Gamma\left({n+1 \over 2}\right) }~.
\eneq
The value of the cosmological constant, $\Lambda$, is given by 
\beeq
\Lambda = { (D-1)(D-2)\over 2 }\lambda~.
\eneq
The effective potential in AdS space is zero at the event horizon ($z\rightarrow -\infty$) and diverges at spatial infinity ($z\rightarrow \eta$, where $\eta$ is a constant discussed in detail in \cite{Cardoso-N-S, Natario-S, Ramin-AdS}\footnote{The symbol used in \cite{Cardoso-N-S} and \cite{Natario-S} for $\eta$ is $x_0$.}).  Therefore, the asymptotic behavior of the solutions is
\beeq
\Psi(z) \approx \left\{ \begin{array}{ll}
                   e^{i\omega z}  & \mbox{as $z \rightarrow -\infty$ }~\\
                   0  & \mbox{as $z\rightarrow \eta$ ~,}
                   \end{array}
           \right.        
\label{asymptotic}
\eneq 
which represents an ``out-going'' wave at the event horizon and no waves at spatial infinity.

To determine the condition on highly real QNMs using the method developed by Cardoso {\it et al}\cite{Cardoso-N-S}, we need to know the behavior of anti-Stokes lines in the complex $r$-plane.  The details of this process have been explained in detail in \cite{Cardoso-N-S, Natario-S}.  As an example, in Fig. \ref{Fig1}, we show the schematic behavior of the anti-Stokes lines in the complex $r$-plane for Schwarzschild-AdS black holes in $D=4$, when $\eta$ is a negative real constant. For large black holes
\beeq
\eta=-{1\over 4 T_H \sin\left({\pi \over D-1}\right)}~,
\eneq
where $T_H$ is the Hawking temperature.
In spacetime dimensions six and higher, we will still have an anti-Stokes line that extends to infinity along the negative real axis and an anti-Stokes line that extends to the event horizon, $R_H$, along the positive real axis.  The only difference is that in higher dimensions we have more anti-Stokes lines extending above and below the real axis.  In order to determine the condition on highly real QNM frequencies, the same paths shown in Fig. \ref{Fig1} will be taken in every spacetime dimension.  For numerically generated pictures of anti-Stokes lines in $D\ge 4$, refer to \cite{Ramin-AdS}.  Once we determine the structure of anti-Stokes lines, we need to know the behavior of the wavefunction $\Psi$ near the singularity at $r=0$ and when $r\rightarrow \infty$.  Near $r=0$, the tortoise coordinate simplifies to
\beeq
z\sim-{r^{D-2}\over 2(D-2)\mu} ~
\label{z0}
\eneq 
and the potential simplifies to
\beeq
V\left[r(z)\right]\sim{j^2-1 \over 4z^2} ~,
\label{v0}
\eneq
where $j=2$ for vector perturbations and $j=0$ for scalar and tensor perturbations\cite{Natario-S}. 
In this region, the differential equation (\ref{Schrodinger}) can be solved exactly.  The solution is 
\beeq
\Psi(z)=A_+ \sqrt{2\pi \omega z}J_{j \over 2}\left({\omega z}\right)+ A_- \sqrt{2\pi \omega z}J_{-{j \over 2}}\left({\omega z}\right)~,
\label{Psi-0}
\eneq 
where $J_\nu$ represents a Bessel function of the first kind and $A_\pm$ are complex constants.
Note that $D$ is an even number because highly real QNMs only exist in even spacetime dimensions\cite{Ramin-AdS}.  Therefore, it is clear that near the origin of the complex plane, $\omega z \ll -1$ on the anti-Stokes line that extends to $-\infty$ along the real axis.  As a result, we can use the approximation
\beeq
J_\nu(x) \sim \sqrt{2\over \pi x}\cos(x+{\nu\pi \over 2}+{\pi \over 4})~~\mbox{when}~x\ll -1~
\label{approx-bessel-1}
\eneq 
to find that
\begin{eqnarray}
\Psi(z)&\sim & 2A_+ \cos(\omega z+\alpha_+)+ 2A_- \cos(\omega z+\alpha_-) \nonumber \\
&=&(A_+ e^{i\alpha_+}+A_- e^{i\alpha_-})e^{i\omega z}+(A_+ e^{-i\alpha_+}+A_- e^{-i\alpha_-})e^{-i\omega z}~.
\label{Psi-small-r}
\end{eqnarray}
Here
\beeq
\alpha_\pm ={\pi \over 4}(1\pm j)~.
\label{alpha}
\eneq 
The result (\ref{Psi-small-r}) is valid on any anti-Stokes line with negative $\omega z$.  Note that in addition to the anti-Stokes line on the negative real axis, $\omega z$ is negative on the anti-Stokes line along the positive real axis.

In the limit where $r \rightarrow \infty$, the tortoise coordinate simplifies to
\beeq
z \rightarrow \eta-{1 \over |\lambda| r}~.
\eneq
In this limit, the QNM potential also takes the simple form 
\beeq
V\left[r(z)\right]\sim{j_\infty^2-1 \over 4(z-\eta)^2} ~,
\label{v-infinity}
\eneq
where $j_\infty=D-1, D-3, D-5$ for tensor, vector, and scalar perturbations respectively.  
Note that the potential (\ref{v-infinity}) is zero for vector (axial) and scalar (polar) perturbations in four spacetime dimensions.  In reality, the QNM potential for these perturbations in four spacetime dimensions asymptotes to a constant as $r \rightarrow \infty$ ($z \rightarrow \eta$).  This may look problematic, but note that we are doing the calculations in the asymptotic limit where $|\omega| \rightarrow \infty$.  Therefore, in the wave equation (\ref{Schrodinger}), replacing the QNM potential, which asymptotes to a constant for large $r$, with a zero should not in principal affect the outcome.  In fact, Natario and Schiappa in \cite{Natario-S} have used exactly the same potential in Eq.\ (\ref{v-infinity}) and their results for polar and axial perturbations in four spacetime dimensions are in perfect agreement with the previous numerical calculations.  It is clear that such modification of the potential at large $r$ does not affect the gap term in the QNM frequency.  However, we suspect that this modification may affect the offset term.  In fact, it is indicated in \cite{Natario-S} that the offset term of polar perturbations does not match the numerical results in \cite{Cardoso-K-L} for large black holes.  This mismatch may originate from the modification of the potential in four spacetime dimensions.  This issue clearly needs further investigation.
The differential equation (\ref{Schrodinger}) can be solved exactly in the region $r\sim \infty$ and the solution is 
\beeq
\Psi(z)\sim B_+ \sqrt{2\pi \omega (z-\eta)}J_{j_\infty \over 2}\left({\omega (z-\eta)}\right)+ B_- \sqrt{2\pi \omega (z-\eta)}J_{-{j_\infty \over 2}}\left({\omega (z-\eta)}\right)~,
\label{Psi-infinity}
\eneq 
where $B_\pm$ are complex constants.
The boundary condition at infinity tells us that the wave function $\Psi$ should die off as $r\rightarrow \infty$.  This requires $B_-=0$.  Also, since  
\beeq
\omega(z-\eta)\sim -{\omega \over |\lambda|r}~
\eneq 
has a positive real value on the anti-Stokes line along the negative real axis, we can use the approximation
\beeq
J_\nu(x) \sim \sqrt{2\over \pi x}\cos(x-{\nu\pi \over 2}-{\pi \over 4})~~\mbox{when}~x\gg 1~
\label{approx-bessel-2}
\eneq 
to write the solution (\ref{Psi-infinity}) as
\beeq
\Psi(z)\sim 2B_+ \cos[\omega (z-\eta)-\beta]
=B_+ e^{-i\beta}e^{i\omega (z-\eta)}+B_+ e^{i\beta}e^{-i\omega (z-\eta)}~,
\label{Psi-infinity-approx}
\eneq 
where
\beeq
\beta ={\pi \over 4}(1 + j_\infty)~.
\label{beta}
\eneq 
Note that the wavefunction $\Psi$ should not change in character along anti-Stokes lines.  Therefore combining Eqs.\ (\ref{Psi-small-r}) and (\ref{Psi-infinity-approx}) results in the first condition on $A_+$ and $A_-$, where
\beeq
\left(A_+ e^{i\alpha_+}+ A_- e^{i\alpha_-}\right)e^{i\omega \eta}e^{i\beta}= \left(A_+ e^{-i\alpha_+}+ A_- e^{-i\alpha_-}\right)e^{-i\omega \eta}e^{-i\beta}~.
\label{Eq1}
\eneq 
In order to impose the boundary condition at the event horizon, we need to move to the anti-Stokes line on the positive real axis which ends at the horizon.  We can do this by a rotation of either $-\pi$ (path 1 in Fig. \ref{Fig1}) or $+\pi$ (path 2 in Fig. \ref{Fig1}) near the singularity at $r=0$. The rotation of $\pm \pi$ in the $r$ coordinate corresponds to a rotation of $\pm\pi(D-1)$ in the tortoise coordinate.  From
\beeq
J_\nu(x)=x^\nu \phi(x)~,
\eneq 
where $\phi$ is an even holomorphic function of $x$, it can be shown that after a rotation of $-\pi(D-1)$ in the tortoise coordinate one can write 
\beeq
A_\pm \sqrt{\omega z}J_{\pm {j\over 2}}(\omega z)\sim e^{-i2(D-2)\alpha_\pm}\cos(\omega z+\alpha_\pm)~~\mbox{when}~\omega z\ll -1~.
\label{-pi-rotation}
\eneq 
Therefore, after a rotation of $-\pi(D-1)$, the solution (\ref{Psi-small-r}) changes to
\begin{eqnarray}
\Psi(z)& \sim & 2A_+ e^{-i2(D-2)\alpha_+}\cos(\omega z+\alpha_+)+ 2A_- e^{-i2(D-2)\alpha_-}\cos(\omega z+\alpha_-) \nonumber \\
  &=&\left(A_+ e^{-i[2(D-2)-1]\alpha_+}+A_- e^{-i[2(D-2)-1]\alpha_-}\right)e^{i\omega z} \nonumber \\
  ~~&+&
\left(A_+ e^{-i[2(D-2)+1]\alpha_+}+A_- e^{-i[2(D-2)+1]\alpha_-}\right)e^{-i\omega z}~.
\end{eqnarray}
According to the boundary condition (\ref{asymptotic}), the solution on the anti-Stokes line ending at the event horizon should behave as $e^{i\omega z}$.  This fact results in a second condition on the coefficients $A_+$ and $A_-$, where
\beeq
A_+ e^{-i[2(D-2)+1]\alpha_+}+A_- e^{-i[2(D-2)+1]\alpha_-}=0~.
\label{Eq2}
\eneq 
Equations (\ref{Eq1}) and (\ref{Eq2}) can only have nontrivial solutions if
\begin{eqnarray}
\left| \begin{array}{cc}
e^{-i[2(D-2)+1]\alpha_+} &e^{-i[2(D-2)+1]\alpha_-} \\
e^{i\alpha_+}e^{i\omega \eta}e^{i\beta}-e^{-i\alpha_+}e^{-i\omega \eta}e^{-i\beta} &e^{i\alpha_-}e^{i\omega \eta}e^{i\beta}-e^{-i\alpha_-}e^{-i\omega \eta}e^{-i\beta}
\end{array}
\right|=0~.
\label{determinant1}
\end{eqnarray}
In four spacetime dimensions, this equation leads to
\begin{eqnarray}
e^{2i\omega \eta}e^{2i\beta}&=&
{e^{-i5\alpha_+}e^{-i\alpha_-}-e^{-i5\alpha_-}e^{-i\alpha_+} \over 
e^{-i5\alpha_+}e^{i\alpha_-}-e^{-i5\alpha_-}e^{i\alpha_+}} \nonumber \\
&=&e^{-i(\alpha_++\alpha_-)}{\sin[2(\alpha_+-\alpha_-)] \over \sin[3(\alpha_+-\alpha_-)]}\nonumber \\
&=& e^{-i(\alpha_++\alpha_-)}{2\cos(\alpha_+-\alpha_-) \over 4\cos^2(\alpha_+-\alpha_-)-1}  ~.
\end{eqnarray}
Since $\alpha_++\alpha_-={\pi \over 2}$ and $\alpha_+-\alpha_-={j \over 2}\pi$, it is easy to show that the above equation leads to 
\beeq
\omega \eta=-\beta-n\pi-{\pi \over 4} +{i\over 2}\ln\left[2\cos\left({j\over 2}\pi\right)-{ 1 \over 2\cos\left({j\over 2}\pi\right)}\right]~~~\mbox{as}~n \rightarrow \infty~.
\label{unphysical-omega}
\eneq 
This result is in perfect agreement with the result obtained by Daghigh and Green in \cite{Ramin-AdS}.  Of course, the sign difference in the imaginary part of $\omega \eta$ is due to the fact that our perturbations depend on time as $e^{i\omega t}$ while Daghigh and Green use $e^{-i\omega t}$.  Since $\eta$ is a negative real constant according to both \cite{Cardoso-N-S} and \cite{Ramin-AdS}, Eq.\ ({\ref{unphysical-omega}) leads to QNM frequencies with negative imaginary part for all types of perturbations.  QNMs with $\omega_I < 0$ cannot be physical because these modes grow with time instead of getting damped.  By repeating the same calculations starting from Eq.\ (\ref{determinant1}), it is easy to show that, in $D\ge 6$ spacetime dimensions, we also obtain QNM frequencies with negative damping which cannot be physical.

Let us now consider path 2 in Fig. \ref{Fig1}, where we have to make a rotation of $\pi$ instead of $-\pi$ in the $r$ coordinate.  We can repeat the steps (\ref{-pi-rotation}) through ({\ref{determinant1}) with a rotation of $\pi (D-1)$ in the tortoise coordinate to find
\begin{eqnarray}
\left| \begin{array}{cc}
e^{i[2(D-2)-1]\alpha_+} &e^{i[2(D-2)-1]\alpha_-} \\
e^{i\alpha_+}e^{i\omega \eta}e^{i\beta}-e^{-i\alpha_+}e^{-i\omega \eta}e^{-i\beta} &e^{i\alpha_-}e^{i\omega \eta}e^{i\beta}-e^{-i\alpha_-}e^{-i\omega \eta}e^{-i\beta}
\end{array}
\right|=0~.
\label{determinant2}
\end{eqnarray}
This equation leads to a condition on $\omega$, where
\begin{eqnarray}
e^{2i\omega \eta}e^{2i\beta}&=&
{e^{i[2(D-2)-1]\alpha_+}e^{-i\alpha_-}-e^{i[2(D-2)-1]\alpha_-}e^{-i\alpha_+} \over 
e^{i[2(D-2)-1]\alpha_+}e^{i\alpha_-}-e^{i[2(D-2)-1]\alpha_-}e^{i\alpha_+}} \nonumber \\
&=&e^{-i(\alpha_++\alpha_-)}{\sin[(D-2)(\alpha_+-\alpha_-)] \over \sin[(D-3)(\alpha_+-\alpha_-)]}\nonumber \\
&=& e^{-i(\alpha_++\alpha_-)}{D-2\over D-3} \cos(\alpha_+-\alpha_-)\nonumber \\ 
&\times& {1-{(D-2)^2-2^2 \over 3!}\sin^2(\alpha_+-\alpha_-)+{[(D-2)^2-2^2][(D-2)^2-4^2] \over 5!}\sin^4(\alpha_+-\alpha_-)-...  \over  
1-{(D-3)^2-1^2 \over 3!}\sin^2(\alpha_+-\alpha_-)+{[(D-3)^2-1^2][(D-3)^2-3^2] \over 5!}\sin^4(\alpha_+-\alpha_-)-... }~.\nonumber \\
\end{eqnarray}
After entering the values for $\alpha_\pm$ , it is easy to show that the above condition on $\omega$ gives
\beeq
\omega \eta=-\beta-n\pi-{\pi \over 4} -{i\over 2}\ln\left[{D-2 \over D-3}\cos\left({j\over 2}\pi\right)\right]~~~\mbox{as}~ n\rightarrow \infty.
\label{physical-omega}
\eneq 
This result again matches perfectly with the result obtained by Daghigh and Green in \cite{Ramin-AdS} considering the difference in the choice of the time dependency of the perturbations.  Since $\eta$ is a negative real constant, it is clear that these QNM frequencies have positive imaginary part.  This indicates the presence of damping which makes these QNMs physically meaningful.  For more arguments on why these modes may be physical see \cite{Ramin-AdS}.

\vskip .5cm

\leftline{\bf Conclusions}
In this paper, we use a different analytic technique to confirm the results obtained by Daghigh and Green in \cite{Ramin-AdS}.  The two analytic methods lead to identical solutions.  The only difference is that the complex coordinate WKB method which is used in \cite{Ramin-AdS} gives two physical and two unphysical solutions, but the technique used in this paper results in only one physical (Eq.\ (\ref{physical-omega})) and one unphysical (Eq.\ (\ref{unphysical-omega})) solution.  No difference has been observed between the outcome of these two analytic methods in the past.  Further investigation of the difference found in this paper may lead to a deeper understanding of these two analytic techniques and their possible limitations.  While this difference is interesting, we do not see it as a major concern.  After all, the calculations are done in the asymptotic limit where the overtone number, $n$, approaches infinity.  In the large $n$ limit, we can neglect all the finite terms in the real part of the highly real QNM frequency.  Neglecting the finite terms will resolve the difference between the two analytic techniques.  The results of this paper leave no doubt that the analytic techniques in the literature\cite{Cardoso-N-S, Ramin-AdS} predict the existence of the highly real QNMs of Schwarzschild-AdS black holes.  If these modes do not exist, the big question is why they appear in analytic calculations which are very successful in producing the correct asymptotic QNM frequencies  of Schwarzschild-AdS black holes.  If these modes do indeed exist, then the question is why the previous numerical calculations\cite{Berti-K, Cardoso-K-L, Konoplya0, high-tone-2} were not able to detect such modes.  It is important to note that numerical calculations should be able to generate the unphysical QNM frequencies as well as the physical ones (see for example \cite{Leaver}).  In order to shed light on these issues, further numerical and analytical investigations are necessary.  

\vskip .5cm

\leftline{\bf Acknowledgments}
We are grateful to Michael Green for numerous useful discussions on the details of these calculations.


\def\jnl#1#2#3#4{{#1}{\bf #2} (#4) #3}

\def\RPP{{\em Reps.\ Prog.\ Phys. }}
\def\Zphys{{\em Z.\ Phys.} }
\def\jssc{{\em J.\ Solid State Chem.} }
\def\jpsJ{{\em J.\ Phys.\ Soc.\ Japan} }
\def\ptps{{\em Prog.\ Theoret.\ Phys.\ Suppl.\ } }
\def\PTP{{\em Prog.\ Theoret.\ Phys.\  }}
\def\LNC{{\em Lett.\ Nuovo.\ Cim.\  }}
\def\LRR{{\em Living \ Rev.\ Relative.} }
\def\JMP{{\em J. Math.\ Phys.} }
\def\NPB{{\em Nucl.\ Phys.} B}
\def\NP{{\em Nucl.\ Phys.} }
\def\PLB{{\em Phys.\ Lett.} B}
\def\PL{{\em Phys.\ Lett.} }
\def\PRL{\em Phys.\ Rev.\ Lett. }
\def\PRB{{\em Phys.\ Rev.} B}
\def\PRD{{\em Phys.\ Rev.} D}
\def\PR{{\em Phys.\ Rev.} }
\def\PRe{{\em Phys.\ Rep.} }
\def\AP{{\em Ann.\ Phys.\ (N.Y.)} }
\def\RMP{{\em Rev.\ Mod.\ Phys.} }
\def\ZPC{{\em Z.\ Phys.} C}
\def\SCI{\em Science}
\def\CMP{\em Comm.\ Math.\ Phys. }
\def\MPLA{{\em Mod.\ Phys.\ Lett.} A}
\def\IJMPB{{\em Int.\ J.\ Mod.\ Phys.} B}
\def\cmp{{\em Com.\ Math.\ Phys.}}
\def\JPA{{\em J.\  Phys.} A}
\def\CQG{\em Class.\ Quant.\ Grav.~}
\def\ATMP{\em Adv.\ Theoret.\ Math.\ Phys.~}
\def\PRSA{{\em Proc.\ Roy.\ Soc.\ Lond.} A }
\def\IJTP{\em Int.\ J.\ Theor.\ Phys.~}
\def\ibid{{\em ibid.} }
\vskip 1cm

\leftline{\bf References}

\renewenvironment{thebibliography}[1]
        {\begin{list}{[$\,$\arabic{enumi}$\,$]}  
        {\usecounter{enumi}\setlength{\parsep}{0pt}
         \setlength{\itemsep}{0pt}  \renewcommand{\baselinestretch}{1.2}
         \settowidth
        {\labelwidth}{#1 ~ ~}\sloppy}}{\end{list}}


\end{document}